\documentclass[12pt,a4paper,onecolumn]{article}
\usepackage{hyperref}
\usepackage[square,numbers]{natbib}
\usepackage{multicol}
\usepackage{graphicx}
\usepackage{mhchem}
\usepackage{amsmath}
\usepackage{physics}
\usepackage{multirow}
\usepackage{slashed}
\usepackage{caption}
\usepackage{mathtools}
\usepackage{color}
\usepackage[top=20mm, bottom=25mm, left=20mm, right=20mm]{geometry}

\begin{document}

\title{Muon Anomalous Magnetic Moment in Noncommutative Space-Time}

\author{Z.\! Rezaei\! \footnote{zahra.rezaei@yazd.ac.ir} \ and A.\! Ghaffary \!\footnote{abozar.gh@gmail.com}\\
Department of Physics, Yazd University\\
\texttt{}
}

\date{\today}

\maketitle 

\begin{abstract}
\maketitle

The explanation of the muon anomalous magnetic moment ($\mu$-AMM) requires new physics beyond the Standard Model. In this work, we investigate the effects of noncommutative space-time on the $\mu$-AMM within the Seiberg-Witten map framework, analyzing both tree-level and loop-level diagrams. Additionally, we examine the $\mu$-AMM by studying the scattering cross-section of electron-positron annihilation into muon-antimuon pairs $ (e^- e^+ \rightarrow \mu^- \mu^+ ) $. Previous studies have explored the implications of noncommutative space-time through various processes, leading to different bounds on the noncommutative parameter $\theta^{\mu\nu}$. Our analysis estimates $\theta^{\mu\nu} \sim  (43 \, \mathrm  TeV)^{-2} $ for both tree-level and loop-level contributions when noncommutative space-time accounts for the entire observed $\mu$-AMM. If noncommutative effects are responsible for only 10$\%$ of the $\mu$-AMM, the constraint relaxes to approximately $(136 \, \mathrm  TeV)^{-2}$. Furthermore, by comparing the noncommutative cross-section of $ e^- e^+ \rightarrow \mu^- \mu^+ $ with experimental data, we derive an approximate bound of $(90 \, \mathrm  TeV)^{-2}$ which corresponds to a 23$\%$ contribution to the $\mu$-AMM.

\end{abstract}

\section{Introduction}
\label{sec.Int}
\leftskip=0.1 cm
\rightskip=0.1 cm

According to electromagnetic theory, any charged particle that rotates possesses a magnetic dipole moment (MDM) and is influenced by the magnetic field. The correlation between the MDM and the characteristics of leptons is demonstrated as: 
\begin{align}
\vec{ \mu } = -g \frac{q}{2m} \vec{ S }
 \label{eq1}
\end{align}
where $\vec{ \mu }$ represents the magnetic moment, and q, m, and $\vec{ S }$ denote the charge, mass and intrinsic angular momentum (spin) of the lepton, respectively. The Dirac equation predicts the g-factor to be precisely 2. However, experimental data from the 1940s revealed that this number is greater than 2 ~\cite{Kusch:1948mvb}. In the same year, quantum electrodynamics (QED) calculations demonstrated that the g-factor must be greater than 2 ~\cite{Schwinger:1948iu}. The deviation of the g-factor from 2 is referred to as the anomalous magnetic moment (AMM) and is given by $a_\mu=\frac{g-2}{2}$, for example for the muon. The Standard Model (SM) theoretical value of AMM($a$) arises from the QED, electroweak (EW) and quantum chromodynamics (QCD) virtual loops calculations 
\begin{align}
a_\mu ^{SM}=a_\mu ^{QED}+a_\mu ^{EW}+a_\mu ^{QCD}.
\end{align}
This equation is a general relation which is written with subscription $\mu$ for the muon, here. Recent compilations provide comprehensive prediction for the SM value of the muon magnetic anomaly \cite{Aoyama:2020ynm,Keshavarzi:2021eqa}. The QED contribution has been computed up to the tenth order in the perturbative expansion, corresponding to O($\alpha^5$)  \cite{Aoyama:2012wk}.  Additionally, the EW contributions have been evaluated up to two-loop precision. These EW SM contributions are given by Feynman diagrams that contain at least one of the EW bosons W, Z, or the Higgs and are strongly suppressed by the heavy masses of the EW bosons \cite{Czarnecki:2002nt,Gnendiger:2013pva}. Hadronic contributions are the most challenging to compute and account for nearly all of the theoretical uncertainty. The dominant hadronic contribution which arises from hadronic vacuum polarization (HVP) and the hadronic light-by-light scattering contribution have been calculated \cite{Kurz:2014wya,Colangelo:2014qya,Hoferichter:2018kwz,Gerardin:2019vio}. 
The muon $g-2$ anomaly is currently given by
\begin{align}
\delta a_\mu= a_\mu^{exp}- a_\mu^{SM}=(24.1\pm 5.4)\times  10^{-10}
\label{eq14/18}
\end{align}
using the latest experimental measurement \cite{Muong-2:2023cdq} and the precise SM prediction computed in \cite{Aoyama:2020ynm,Keshavarzi:2021eqa}. Recent lattice field theory studies \cite{Ce:2022kxy,Borsanyi:2020mff} have cast doubt on  Beyond Standard Models (BSM) explanations for this anomaly. If Eq. \ref{eq14/18} is considered valid, it indicates a significant discrepancy of approximately $4.5 \sigma$ \cite{Wang:2023njt} from the SM predictions. The significant gap between theoretical predictions and experimental results suggests the existence of New Physics (NP) beyond the SM \cite{Wang:2023njt}. \\
 Particle physicists are actively exploring BSM, also referred to as  NP, in order to elucidate the $\mu$-AMM,
\begin{align}
a_\mu ^{exp}=a_\mu ^{SM}+a_\mu ^{NP}.
\end{align}

BSM encompasses two different approaches. The initial approach, exemplified by supersymmetry \cite{Nilles:1983ge,Dong:2024lvs,Jim:2024etp}, dark matter (DM) \cite{Arkani-Hamed:1998sfv,Rezaei:2020mqj,Saha:2024dhr} and two Higgs doublet model (2HDM) \cite{Branco:2011iw,Liu:2023gwu}, involve the addition of new particles and interactions to SM. However, a significant drawback of these theories is that the anticipated new particles have yet to be detected. On the other hand, the second approach contains theories that introduce new SM interactions.  The Lorentz Violation (LV)  \cite{Colladay:1998fq,Haghighat:2013xja,Sis:2023emv,Ebrahimi:2025hjx} and the noncommutative Space-Time (NCST) theories \cite{Douglas:2001ba,Rezaei:2018mlk}, which established a lower bound for string theory \cite{Chu:1998qz}, are the most famous of these theories.


Due to the SM's limitation in explaining the inconsistency between the measured and predicted values of the muon MDM, we investigate this anomaly by examining NCST as a potential NP framework beyond the SM. 

The theoretical calculations in NCST are divided into two parts: the U* approach \cite{Chaichian:2001py} and the Seiberg-Witten map (SWM) \cite{Seiberg:1999vs}. NCST adjusts existing vertices and allows new couplings. Triple gauge boson couplings ($\gamma\gamma\gamma$, $\gamma\gamma$ZZ, $\gamma$ZZ, etc.) result from changes in space-time behavior. Numerous studies have been carried out on the muon dipole moment in both approches.  Here, we review some of them and introduce our calculations based on SWM.

The U* approach to NCST is primarily associated with the study of NC geometry and its implications for quantum field theories, particularly in the context of Yang-Mills theories. This approach modifies the traditional commutative algebra of functions on space-time by introducing a NC structure, which fundamentally alters the way we understand the geometry of space-time. This approach employs a novel phase-space analysis of dynamical fields. This allows for a systematic treatment of gauge theories in NC spaces, where the usual gauge invariance is modified due to the noncommutativity of coordinates \cite{Ahmadiniaz:2018olx}. Therefore, it has potential applications in various areas, including quantum gravity and string theory, where understanding the fundamental structure of space-time is crucial. It provides insights into how particles behave at high energies and small distances, where classical descriptions fail ~\cite{Mendes:2016jhj,Majid:2006xn}.

 The contribution of the U* approach to the AMM is evaluated up to the loop level. In Ref.~\cite{Riad:2000vy}, it has been shown that the tree level portion of magnetic moment is zero and an intrinsic spin-independent magnetic moment arising from the noncommutative structure in the one loop level is a unique signature of NCST. On the other hand, the contribution of the EW one loop level is investigated in Ref.~\cite{Kersting:2001zz}. It has been showen that if the scale of noncommutativity is of order O(TeV), the shift in $\mu$-AMM due to the altered structure can be  as large as one part in $10^{8}$ which is larger than the amount in dispute. Additionally, in Ref.~\cite{Haghighat:2014nra}, the $\mu$-AMM is investigated with respect to the muonic hydrogen lamb shift limits($\Lambda_{NC}=1.5$ TeV):
\begin{align}
\delta a_\mu  =\frac{5}{6} \frac{\alpha}{2 \pi} \frac{p^2}{\Lambda^2} \sim 3 \times 10^{-9}.
\label{eq14.23}
\end{align}
However, it is necessary to reassess these values since each component should be smaller than the threshold value (Eq.(\ref{eq14/18})).

In SM, the electron positron scattering experiments play a crucial role in resolving $\mu$-AMM by improving the HVP contribution, which dominates the theoretical uncertainty. In addition, due to the discrepancy between the experimental data of $ (e^- e^+ \rightarrow  l^-  l^+ ) $ cross-section and the SM prediction, we examine the NC cross-section and obtain its effect on the $\mu$-AMM. We compute the cross-section of $ e^- e^+ \rightarrow \mu^- \mu^+ $ utilizing the SWM,  referencing the SM cross-section as a guide. Given that the SWM correction is scalar, we extend our calculations to five loops.  Ultimately, based on the cross-section data  $ (e^- e^+ \rightarrow \mu^- \mu^+ ) $, it is possible to determine the NC parameter. Upon establishing the NC parameter, we can forecast the AMM of the muon.

The structure of article is as follows: the NCST is briefly described in Sec.~\ref{sec:NCST}, where we also present the NC parameter and discuss some of its constraints. The calculation of AMM has been conducted at both the tree and loop levels within the framework outlined in Sec.~\ref{sec:AMM in NCST}. In Sec.~\ref{Scattering}, we derive the cross section of $ (e^- e^+ \rightarrow \mu^- \mu^+ ) $ in NCST.  We give a brife review on the $\mu$-AMM in NCST and explain our results in Sec.~\ref{sec:results}. Concluding remarks and also future points of view are summarized in  Sec.~\ref{sec:conc}.


\section{Noncommutative Space-Time (NCST) and NC parameter}\label{sec:NCST}

Heisenberg initially proposed the concept of NC, which was later published by Snyder \cite{Snyder:1946qz}. This concept aimed to provide an explanation for the infinities encountered in QFT. However, with the advent of the renormalization theory, these issues were resolved, leading to a decline in the acceptance of the NC theory. Nonetheless, NC theory regained prominence in the 1980s–1990s with the rise of string theory, particularly through studies of strings displaying noncommutative properties \cite{Seiberg:1999vs,Witten:1985cc}. \\
In NC theory, the space-time coordinates do not commute with each other,
\begin{align}
[x^\alpha , x^\beta]=\theta ^{\alpha \beta}
\end{align}

where  $x^\alpha$ ,  $x^\beta$  denote the space-time 4-vectors and $\theta^{\alpha \beta}$ is the antisymmetric tensor with constant and real value, $|\theta| = \frac{1}{\Lambda_{NC}^2}$, where  $\Lambda_{NC}$ is the energy scale of the NC theory.  The NC parameter $\theta^{\alpha \beta}$ (or equivallently NC energy scale $\Lambda_{NC}$) plays a crucial role in NC theory.  

Numerous studies have been carried out on NC theory in different experiments. Based on these investigations, $\Lambda_{NC}$ exhibits a broad spectrum, ranging from several 10 GeV to several 100 TeV ~\cite{Riad:2000vy,Kersting:2001zz,Schupp:2002up,Wang:2001ig,Rafiei:2016lqq,Rezaei:2018nrv,Rafiei:2022tsd,Horvat:2010sr,Adorno:2012jh}. 

These constraints are derived from experimental and observational data across various fields. Variations in the numerical values of NC across different models with similar vertices arise from differences in the experiments used during analysis. One possible classification of these experiments is presented in the following:  $\textbf{i}$) Particle Physics (Collider Experiments): $\textbf{a}$) searches for deviations from SM predictions (e.g., anomalous particle scattering, Lorentz violation in Large Hadron Collider) constrain the NC scale to $\Lambda_{NC}>$ O(1–10 TeV) and  $\textbf{b}$) earlier collider data like LEP and Tevatron ruled out low-scale noncommutativity ($< $100 GeV) \cite{OPAL:2003eoc}, $\textbf{ii}$) Astrophysics, Cosmic Rays and Gravitational Waves searches suggest that the relevant energy scale could be less than the Planck mass \cite{Biller:1998hg},  $\textbf{iii}$) quantum optics and precision tests: precision measurements of atomic energy levels and lepton magnetic moments constrain NC effects to scales $\Lambda_{NC}>$O(1–10 TeV) \cite{Chaichian:2000si,Alavi:2005yr}. Furthermore, some solid-state systems simulate noncommutative geometry which are low-energy analogs and don’t directly constrain fundamental NCST. Next-generation colliders (Future Circular Collider-FCC), CMB-S4, and quantum gravity experiments  could strengthen bounds.

As mentioned in Sec.~\ref{sec.Int}, there are two approaches to explain NCST. Following this, we initiate our analysis by computing the $\mu$-AMM at the tree level and loop level through the application of the SWM.
 
\begin{center}
\includegraphics[height = 2cm]{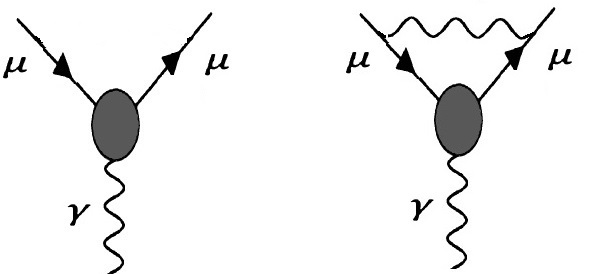} 
\captionof{figure}{Tree level and 1-loop level correction in NCST}
\label{fig-treelevel}
\end{center}

\section{Contribution of AMM in NCST}
\label{sec:AMM in NCST}

The corrected photon-fermion vertex  in NCST, Fig. \ref{fig-treelevel}, is given by \cite{Melic:2005fm}:
\begin{equation}
ie \Gamma^\alpha = ieQ_f \gamma^\alpha +\frac {1}{2} eQ_f [(p_{2}  \theta  p_{1})\gamma^\alpha  -(p_{2}  \theta)^\alpha (\slashed{p}_{1} - m_f) -(\slashed{p}_{2} - m_f) (p_{1}  \theta) ^\alpha]
\label {eq 10.8}
\end{equation}
where $Q_f$ and $m$ are the fermion's charge and mass and $p_{1}$ and $p_{2}$ show the input and output fermion's momentum, respectively. Figure \ref{fig-treelevel} (left part) displays that the NC theory diagram is identical to the SM tree-level diagram, except for the NC correction in the vertex.  

In the absence of NC theory, the tree level amplitude is: 
\begin{equation}\label{eq:SM}
iM=-\bar{u}^{(s)} ie \gamma^\alpha {u}^{(s^\prime)} A _\alpha(q),
\end{equation}
using the Gordon identity \cite{Crivellin:2022idw},
\begin{align}
iM=\frac{ie}{2m} \bar{u}^{(s)}(p^\alpha   A _\alpha(q)  -  \frac{1}{2} \sigma_ {\alpha \beta} F^{\alpha \beta}(q)) {u}^{(s^\prime)},
\label {eq 10.11}
\end{align}
where the initial expression pertains to the electrical interaction while the subsequent expression represents the magnetic interaction. Our focus is on the AMM, which is influenced by magnetic interactions, therefore we disregard the electrical component \cite{Crivellin:2022idw}. 
By replacing the corrected vertex $ie \Gamma^\alpha$ (Eq.~(\ref{eq 10.8})) insted of $ie \gamma^\alpha$ into the amplitude Eq.~(\ref{eq:SM}), it could be written as:
\begin{align}
iM=i(M^{'} + \delta M^{'}),
\end{align}
where the SM amplitude contribution, $M^{'}$\!\!, is:
\begin{equation}\label{eq:Mprim}
M^{'}=\frac{e}{4m} \bar{u} \, \sigma_ {\alpha \beta} \, {u} \,  F^{\alpha \beta}(q)= \frac{e}{m} B S^z ,
\end{equation}
 The right term is obtained introducing the muon's charge, mass, and spin (represented by e, m and $S^z$, respectively) into a magnetic field aligned with the Z-axis.  
The pure NC  amplitude,  $\delta M^{'}$\!\!, defines as:
\begin{align}
\delta M^{'}=-\bar{u}^{(s)} ie  \gamma^{\alpha}  {u}^{(s^\prime)} A _\alpha(q)=-\bar{u}^{(s)}{eQ_f}\gamma^\alpha (\frac{p_{2}  \theta  p_{1}}{2}) {u}^{(s^\prime)} A _\alpha(q),
\label{eq 15.15}
\end{align}
the last two terms of Eq.~(\ref{eq 10.8}) have been excluded due to the Dirac equation, leading to the formulation of Eq.~(\ref{eq 15.15}). Consequently, by applying the Gordon identity, $\delta M^{'}$ at the tree level is defined as follows:
\begin{align}
\delta M^{'}= \frac{e}{2m} (p_{2}  \theta  p_{1}) B S^z .
\end{align}
 As a result, the NCST tree-level calculation of the AMM predicts 
\begin{align}
\delta a_\mu ^{tree-level} =\frac{m}{e} \frac{\delta M{'}}{B S^z} = \frac{p_{2}  \theta  p_{1}}{2} .
\label{eq14}
\end{align}
Since the implementation of the SWM approach, the calculation has remained incomplete, therefore, our goal is to determine the contribution at the loop level. The calculation of 1-loop vertex modifications as shown in Fig.~\ref{fig-treelevel} (right part), for the $\mu$-AMM resulted in
\begin{align}
\delta a_\mu ^{1-loop} =\frac{\alpha_e}{2 \pi} \frac{p_{2}  \theta  p_{1}}{2} 
\label{eq14.5}
\end{align}
which reffered to the Schwinger 1-loop modification \cite{Mandl}, where $\alpha_e$ is fine structure constant.

Given that $\frac {1}{2} (p_{2}  \theta  p_{1})$ is a scalar quantity, it can be regarded as a factor in the computation of vertex modification. The contribution of 5-loop is calculated \cite{Aoyama:2012wk} in SM. Our objective is to expand the scope of these computations in NCST loop level.
\[
\delta a_{\mu} = (\frac {p_{2}  \theta   p_1}{2}) \Big \{  1+(a^{1-loop}_{\mu})_{SM}+ (a^{2-loop}_{\mu})_{SM}
\]
\vspace{-0.6cm}
\begin{align}
+ (a^{3-loop}_{\mu})_{SM}+ (a^{4-loop}_{\mu})_{SM} + (a^{5-loop}_{\mu})_{SM} \Big\}.
\label {eq 22/1} 
\end{align}
The initial term represents the contribution from the tree level, while the subsequent terms account for the contributions at the loop level. Subsequently, we will utilize the data provided in Table \ref{tab 1.1} \cite{Aoyama:2012wk} to summarize the $\mu$-AMM equation,
\begin{align}
\delta a_{\mu} = (\frac {p_{2}  \theta   p_1}{2})  [1+(\frac {\alpha_e}{2 \pi}) +...],
\label {eq 22/8} 
\end{align}
which is the first tree and loop level calculations of  $\mu$-AMM in SWM. In fact, the vertex modification in SWM has been corrected as follows:
\begin{align}
\Gamma^\alpha = \gamma^\alpha \Big \{ 1+(\frac {p_{2}  \theta   p_1}{2}) [1+(\frac {\alpha_e}{2 \pi}) +...]\Big \}.
\label {eq 22/13} 
\end{align}
We will discusse the results in Sec.~\ref{sec:results}. In the following section, we will examine the impact of NC on the scattering experiments.

\begin{center}
\captionof{table}{Contributions to muon g-2 from QED perturbation term $a_\mu ^{(2n)} (\frac{\alpha}{ \pi})^n \times 10^{11} $ ~\cite{Aoyama:2012wk}.} \label{tab 1.1}
\begin{tabular}{|p{1.8cm}|p{3.5cm}|p{3.5cm}|}
 \multicolumn{3}{c}{} \\
  \hline
   \textbf{order} & \textbf{ with $\alpha^{-1}$ (Rb)}& \textbf{ with $\alpha^{-1} (a_e)$} \\
\hline
\textbf{$ \quad 2 $} & $ 116 140 973.318 (77) $ & $  116 140 973.212 (30) $\\
  \hline
\textbf{$  \quad 4 $} & $ 413 217.6291 (90)$ & $  413 217.6284 (89)$\\
  \hline
\textbf{$ \quad 6 $} & $30 141.902 48 (41) $ & $  30 141.902 39 (40) $\\
  \hline
\textbf{$ \quad 8 $} & $381.008 (19) $ & $  381.008 (19) $\\
  \hline
\textbf{$ \quad 10 $} & $ 5.0938 (70) $ & $ 5.0938 (70) $\\
  \hline
\textbf{$ a_\mu(QED) $} & $116 584 718.951 (80) $ & $  116 584 718.845 (37) $\\
  \hline
 \multicolumn{3}{c}{} \\
\end{tabular}
\end{center}

\begin{center}
\includegraphics[height=1.5 cm]{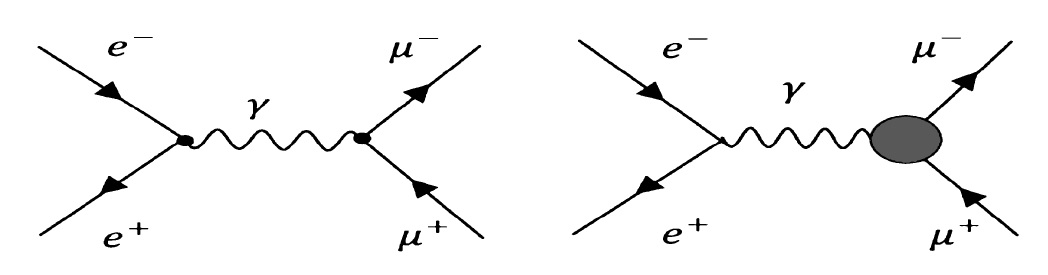} 
\captionof{figure}{electron positron scattering in NC theory}
\label{fig2.2}
\end{center}

\section{ Scattering in NCST}\label{Scattering}

Based on Fig.~\ref{fig2.2}, the cross-sectional view of $ (e^- e^+ \rightarrow \mu^- \mu^+ ) $ in the SM (on the left) and the NC scenario (on the right) is illustrated. We ignore the contribution of $Z$ boson mediatted contribution because its effect is so small in our calculation.

The cross-section in the SM is denoted by
\begin{align}
\left(\frac{d \sigma}{d\Omega}\right)_{CoM} =\frac{1}{64 \pi^2 (E_1+E_2)^2} \frac{|p^{'}_1|}{|p_1|}(\prod_{l} 2m_l)| \mathcal{M}|^2 ,
\end{align}
where the scattering amplitude  is
\begin{eqnarray}
-i\mathcal{M}_{SM}(e^- e^+ \rightarrow \mu^- \mu^+ ) = [\bar{v}(p_2) ie\gamma^\mu u(p_1)] 
              (\frac{-i g_{\mu\nu}}{s^2}) [\bar{u}(p_3) ie a \gamma_\mu v(p_4)],
\label{amplitude}
\end{eqnarray}
hence, the cross-section in SM obtains \cite{Mandl},
\begin{align}
\sigma_{SM} = \frac{4 \pi \alpha^2}{3s}.
\label{eq 21.9}
\end{align}
The contribution of NC in scattering is represented by  
\begin{align}
\mathcal{M}=\frac{-e^2}{s} [\bar{v}(p_2) \gamma^\mu u(p_1)] [\bar{u}(p_3) \gamma_\mu  \Big \{ 1+(\frac {p_{2}  \theta   p_1}{2}) [1+(\frac {\alpha_e}{2 \pi}) +...]\Big \} v(p_4)],
\label{22.7}
\end{align}
where the first bracket is the SM contribution ($ [\bar{v}(p_2) \gamma^\mu u(p_1)]$) and the second bracket contains the contribution of NCST. It is evident that the corrected vertex (Eq.~\ref{eq 22/13}) has been utilized here. Giving the $\theta =0$, the NC amplitude (Eq.~(\ref{22.7})) decreases to the SM one (Eq.~(\ref{amplitude})). The second term in braket $\{\}$, i.e. $(\frac {p_{2}  \theta   p_1}{2}) [ 1+(\frac {\alpha_e}{2 \pi}) +...]$, as a scalar, contributes only as a coefficient within the amplitude’s expression. Hence, Eq.~(\ref{22.7}) is reformulated as:
\begin{align}
\mathcal{M}=\mathcal{M}_{SM}+\Delta \mathcal{M}_{SM},
\end{align}
where $\Delta = \Big \{ (\frac {p_{2}  \theta   p_1}{2}) [1+(\frac {\alpha_e}{2 \pi}) +...] \Big \}$.
Therefore, the square of $\mathcal{M}$ is represented by:
\begin{align}
\mathcal{M}^2=\mathcal{M}^2_{SM}(1+2\Delta),
\end{align}
where the impact of $\Delta^2$ is negligible because of its small magnitude. Finally, the cross-section in the presence of NC effects is:
\begin{align}
\sigma_{NC} = \frac{4 \pi \alpha^2}{3s}\Big \{ p_{2}  \theta   p_1 [1+(\frac {\alpha_e}{2 \pi}) +...] \Big \}
\label{26.5}
\end{align}
In the following, the NC parameter is determined using the experimental cross-section measurments and results are discussed.

\section{results and discussion}\label{sec:results}
Now let us examine the previous outcomes that were achieved using a random NC parameter value.

The initial approach, referred to as U*, for a muon energy of approximately 3GeV and $\theta \sim 1.36 \times 10^7 (\mathrm{MeV})^{-2}$ or $(\mathrm{0.27 KeV})^{-2}$, forecasts  $\delta a_\mu= 1.57 \cross 10^{-9}$ at the loop level, taking into account the 3-photon vertex contribution ~\cite{Wang:2001ig}.

The additional loop level QED contribution in this approach yielded $\delta a_\mu \sim  3\cross 10^{-9}$ when $p \sim 3 \mathrm{GeV}$ and $\theta \sim (\mathrm{1.5 TeV})^{-2}$  ~\cite{Haghighat:2014nra}.

Ultimately, the expected $\delta a_\mu$ for the EW loop level contribution involving the W boson is about $10^{-8}$ when $p \sim 3 $ GeV and $\theta \sim(1 \mathrm{TeV})^{-2}$ ~\cite{Kersting:2001zz}. 

The U* approach indicates a variation in the range of NC parameter from $(\mathrm{KeV})^{-2}$ to $(\mathrm{TeV})^{-2}$ in order to account for the AMM of the muon, despite the expected constancy of the NC parameter.

It is understood that the combined contributions from QED, EW theory and QCD must account for the discrepancy presented in Eq.~(\ref{eq14/18}). However, the 1-loop NCQED contribution in U* aproach, mentioned above, surpasses this deviation, requiring the results to be re-evaluated.

 As previously indicated, the second NC theory method is SWM. At this point, we aim to investigate scattering experiments and propose that NC theory explains the differences noted between theoretical predictions and experimental findings. The calculation of the $\mu$-AMM contribution is presented in Sec. \ref{sec:AMM in NCST},
\begin{align}
\delta a_{\mu} = (\frac {p_{2}  \theta   p_1}{2})  [1+(\frac {\alpha_e}{2 \pi}) +...].
\label {eq 22/22} 
\end{align}
Given a muon energy of  3 GeV~\cite{Haghighat:2014nra} and  $\mu$-AMM (Eq.~(\ref{eq14/18})), we estimate the NC parameter to be approximately $(43 \, \mathrm TeV)^{-2}$ within the framework of the SWM. This value fully accounts for the observed discrepancy. Since such a discrepancy may arise from multiple contributions, we consider the scenario in which $10 \%$ of this difference originates from NCST effects.  Using Eq.~(\ref{eq14/18}) with  $p \sim 3 $ GeV, this assumption constrains the NC parameter to $(136 \, \mathrm TeV)^{-2}$.

Given that the storage ring operates with a maximum energy of 3 GeV, we focus our investigation on performing scattering tests at higher energy scales where this constraint does not apply. Based on Eq.~(\ref{26.5}), the NC cross-section's independence from energy becomes apparent (with $p_1=p_2=\frac{\sqrt{s}}{2}$). Therefore, it is crucial to determine the region in which the correlation between the SM cross-section and energy is minimized. According to Eq.~(\ref{eq 21.9}), this occurs exclusively in the high-energy region.  This may explain why, in scattering experiments, it is anticipated that the effects of NC theory manifest at high energy levels. 

According to Eqs.~(\ref{eq 21.9}) and (\ref{26.5}), the ratio of the NC cross-section to that of the SM is defined as:
\begin{align}
|\frac{\Delta \sigma}{\sigma_{SM}}| \sim |\frac{\sigma_{NC}}{\sigma_{SM}}|= p_{2}  \theta   p_1 [1+(\frac {\alpha_e}{2 \pi}) +...].
\label{delta5}
\end{align}
It is now necessary to determine the NC parameter. The ratio of the $\Delta \sigma=\sigma_{EXP}-\sigma_{SM}$ cross-section to the SM cross-section is approximately $1.26 \times 10^{-4}$  in the high-energy region, as shown in Table \ref{tab 2}. We consider the NC theory accounts for the observed discrepancies across 1–10 TeV muon energies, consistent with the data in Table  \ref{tab 1.1}. Finally, Eq.~(\ref{delta5}) yields values of the NC parameter ranging from  $(90 \, \mathrm TeV)^{-2}$ at 1 TeV to ($900 \, \mathrm TeV)^{-2}$ at 10 TeV. 


While high-energy experiments, such as scattering processes, are expected to reveal NC effects, there also exists the possibility of detecting such effects in low-energy experiments. These include precision measurements of decay rates or Resonance Spectroscopy (MRS)-based observations, potentially enhanced by advanced atomic and optical clock techniques \cite{Safronova:2017xyt}. It is noteworthy that the NC parameter was estimated to be $(240 \, \mathrm TeV)^{-2}$ using the MRS method \cite{Adorno:2012jh}. 

In conclusion, by referencing Eq.~(\ref{eq14/18}) and utilizing the parameter values $ \theta \sim (90 \, \mathrm{TeV})^{-2}$ and $p = 3 $ GeV, the SWM modification (Eq.~\ref{eq 22/22}) yields:
\begin{align}
\delta a_{\mu}= 5.56 \times 10^{-10}, 
\label{eq19/18}
\end{align}
which describes approximatly 23 $\%$ of the $\mu$-AMM of Eq.~(\ref{eq14/18}).  It is important to emphasize that this prediction stems from the NCQED contribution in the tree-level calculations and extends up to the fifth-loop order.

\begin{center}
\captionof{table}{The ratio of scattering in High energy levels \cite{ParticleDataGroup:2022pth}} \label{tab 2}
\begin{tabular}{p{1.4cm}|p{4 cm}} 
  \hline
 \textbf{Energy} & \textbf{$ \quad |\frac{\sigma_{EXP}-\sigma_{SM}}{\sigma_{SM}}|$} \\
  \hline
 1 TeV &  \textbf{\quad $ 1.2593789 \times 10^{-4} $} \\
 \hline
 10 TeV &  \textbf{\quad $ 1.259379627 \times 10^{-4} $} \\
  \hline
 \multicolumn{2}{c}{} \\

\end{tabular}
\end{center}

\section{Conclusion}\label{sec:conc}

The $\mu$-AMM is remained one of the unresolved issues within the SM, despite the accurate computations of contributions from QED, EW and QCD. While numerous models have been proposed to account for this anomaly, we focus on the NC theory in this manuscript.

In the U* approach of NCST,  $\mu$-AMM is predicted in a  single loop, exceeding anticipated results and thus requiring adjustment while the range of the NC parameter varies from $( \mathrm KeV)^{-2}$ to $( \mathrm TeV)^{-2}$. 

Within the SWM framework, we have carried out vertex modification calculations up to five loops which has not been before accomplished.  
When attributing the full $\mu$-AMM to NCST-derived effects, the noncommutativity parameter is constrained to $(43 \, \mathrm TeV)^{-2}$. In a scenario where only $10 \%$ of the $\mu$-AMM stems from NCST, the parameter estimate relaxes to $(136 \, \mathrm TeV)^{-2}$, aligning with the scale probed by MRS tests.

To probe high-energy effects, we have analyzed the $\mu$-AMM  through precision measurements of $e^- e^+ \rightarrow \mu^- \mu^+ $ scattering processes.  We constrained the NC parameter by associating the $\Delta \sigma$ difference with NCST effects and applied the resulting value ($\theta \sim (90 \mathrm TeV)^{-2}$) to determine the $\mu$-AMM contribution. The $\delta a_{\mu}= 5.55 \times 10^{-10}$ describes approximatly 23 $\%$ of $\mu$-AMM. Future colliders (e.g., FCC-ee, ILC, CLIC) could probe these effects through precision measurements of differential cross-sections (also angular distributions).

\end{document}